\documentclass[fleqn,11pt,twocolumn]{article}

\usepackage[dvips]{graphicx}
\title{%
\hfill
{\small DPNU-05-10 }
\\
{\bf Pion Velocity near the Chiral Phase Transition Point }
\\
{\bf in the Vector Manifestation }~\footnote{
  Contribution to the proceedings of the YITP workshop on 
  Non-equilibrium dynamics in the QCD phase transitions,
  February 22 - 24, 2005, YITP, Kyoto, Japan.
}
\\
\vspace{5mm} \hfill \normalsize Masayasu Harada~$^{(a)}$, Mannque
Rho~$^{(b)}$
and Chihiro Sasaki~$^{(c)}$\\
\hfill {\it {\rm (a)}~Department of Physics, Nagoya University,
Nagoya, 464-8602, Japan,
}\\
\hfill {\it {\rm (b)}~Service de Physique Th\'eorique, CEA Saclay,
91191 Gif-sur-Yvette, France,
}\\
\hfill {\it {\rm (c)}~Gesellschaft f\"{u}r Schwerionenforschung
(GSI), Planckstr. 1, 64291 Darmstadt, Germany } }
\date{\empty}

\begin{document}

\maketitle
\begin{abstract}
We study the pion velocity near the critical temperature $T_c$ of
chiral symmetry restoration in QCD. Using chiral perturbation
theory based on hidden local symmetry (HLS) as an effective field
theory, where the chiral symmetry restoration is realized as the
vector manifestation (VM), we show that the pion velocity for $T
\to T_c$ is close to the speed of light, $v_\pi (T) = 0.83 - 0.99$
which is at variance with the sigma model result which predicts
$v_\pi (T_c) = 0$. Our prediction on $v_\pi (T_c)$ is compared
with the value extracted from the recent RHIC data.

\end{abstract}


\section{Introduction}
It has been suggested by Cramer et al.~\cite{Cramer} that the
recent result by the STAR collaboration at RHIC~\cite{star}
provides information on the pion velocity near the chiral phase
transition in hot medium with the conclusion that the pions seen
in Hanbury Brown-Twist interferometry are emitted from a
chiral-symmetry restored phase~\cite{Wilczek}. In this note, we
wish to interpret the result of the analysis by Cramer et al. in
terms of the vector manifestation scenario of hidden local
symmetry theory~\cite{HY:VM,HY:PR}.

The effective field theory based on the 
hidden local symmetry (HLS),
which includes both pions and vector mesons as the
dynamical degrees of freedom, implemented with the Wilsonian
matching to determine the bare theory from the underlying QCD,
leads to the vector manifestation (VM) of chiral
symmetry~\cite{HY:VM,HY:PR} in which the massless vector meson becomes
the chiral partner of the pion at the critical point~\footnote{ As
studied in Ref.~\cite{HY:PR} in detail, the VM is defined only as
a limit with bare parameters  approaching the VM fixed point from
the broken phase.}. This picture provides a strong support for
Brown-Rho scaling~\cite{BR} which predicted that the mass of
light-quark hadrons should drop in proportion to the quark
condensate $\langle\bar{q}q\rangle$. By now there are several
experimental indications that this scenario is a viable one. The
earliest one was the enhancement of dielectron mass spectra below
the $\rho / \omega$ resonance observed at CERN SPS~\cite{ceres}
which has been satisfactorily explained by the dropping of the
$\rho$ meson mass~\cite{LKB} according to the Brown-Rho
scaling~\cite{BR}. This explanation however is not unique as there
are alternative -- but not necessarily unrelated -- mechanisms
that can equally well describe the presently available
data~\cite{rapp-wambach}. A much more compelling evidence comes
from the mass shift of the $\omega$ meson in nuclei measured by
the KEK-PS E325 Experiment~\cite{KEK-PS} and
the CBELSA/TAPS Collaboration~\cite{trnka}, and also from that of
the $\rho$ meson observed in the STAR experiment~\cite{SB:STAR}.
These are clean signals manifested in a ``pristine" environment
unencumbered by a plethora of ``trivial" effects.

In this note, we focus on the pion velocity near the critical
temperature and make a prediction based on the vector
manifestation (VM). The pion velocity is one of the important
observable quantities in heavy-ion collisions as it controls the
pion propagation in medium through a dispersion relation. Our
prediction for $v_\pi$ is $v_\pi (T_c)= 0.83 - 0.99$, which we
should stress, is at strong variance with the result obtained in
sigma models~\footnote{By sigma models, we mean generically the
chiral symmetry models that contain pions as the $only$ relevant
long-wavelength degrees of freedom.}, i.e., $v_\pi (T_c)=
0$~\cite{SS}. We believe our result to be consistent with $v_\pi
(T) = 0.65$ of Cramer {\it et al.}~\cite{Cramer} extracted from
the recent STAR data~\cite{star}. What distinguishes our approach
from that of sigma models is the {\it intrinsic temperature and/or
density dependence of the parameters} of the HLS Lagrangian, that
results from integrating out the high energy modes (i.e., the
quarks and gluons above the matching scale) {\it in
medium}~\cite{VM:hot,VM:dense}. It is this intrinsic temperature
and/or density dependence -- which causes Lorentz symmetry
breaking -- that plays the essential role for realizing the chiral
symmetry restoration in a consistent way and underlies the
Brown-Rho scaling.


\section{Vector manifestation of chiral symmetry}
In this section, we start with the HLS Lagrangian at leading order
including the effects of Lorentz non-invariance. Then we present
the conditions satisfied at the critical point.

The HLS theory is based on
the $G_{\rm{global}} \times H_{\rm{local}}$ symmetry,
where $G=SU(N_f)_{\rm L} \times SU(N_f)_{\rm R}$ is 
the chiral symmetry
and $H=SU(N_f)_{\rm V}$ is the HLS.
The basic quantities are
the HLS gauge boson $V_\mu$ and two matrix valued
variables $\xi_{\rm L}(x)$ and $\xi_{\rm R}(x)$
which transform as
$\xi_{\rm L,R}(x) \to \xi^{\prime}_{\rm L,R}(x)
  =h(x)\xi_{\rm L,R}(x)g^{\dagger}_{\rm L,R}$,
where $h(x)\in H_{\rm{local}}\ \mbox{and}\ g_{\rm L,R}\in
[SU (N_f)_{\rm L,R}]_{\rm{global}}$.
These variables are parameterized as
\footnote{
 The wave function renormalization constant of the pion field
 is given by the temporal component of the pion decay 
 constant~\cite{sasaki,Meissner:2001gz}.
 Thus we normalize $\pi$ and $\sigma$ by $F_\pi^t$ and $F_\sigma^t$
 respectively.
}
$ \xi_{\rm L,R}(x)=e^{i\sigma (x)/{F_\sigma^t}}
     e^{\mp i\pi (x)/{F_\pi^t}}$,
where $\pi = \pi^a T_a$ denotes the pseudoscalar Nambu-Goldstone
(NG) bosons associated with the spontaneous symmetry breaking of
$G_{\rm{global}}$ chiral symmetry, and $\sigma = \sigma^a T_a$
denotes the NG bosons associated with the spontaneous
breaking of $H_{\rm{local}}$. This $\sigma$ is absorbed into the
HLS gauge boson through the Higgs mechanism, and then the vector
meson acquires its mass. $F_\pi^t$ and $F_\sigma^t$ denote
the temporal components of the decay constant of
$\pi$ and $\sigma$, respectively.
The covariant derivative of $\xi_{L}$ is given
by
\begin{equation}
 D_\mu \xi_L = \partial_\mu\xi_L - iV_\mu \xi_L
 {}+ i\xi_L{\mathcal{L}}_\mu,
\end{equation}
and the covariant derivative of $\xi_R$ is obtained
by the replacement of ${\mathcal L}_\mu$ with ${\mathcal R}_\mu$
in the above where
$V_\mu$ is the gauge field of $H_{\rm{local}}$, and
${\mathcal{L}}_\mu$ and ${\mathcal{R}}_\mu$ are the external
gauge fields introduced by gauging $G_{\rm{global}}$ symmetry.
In terms of ${\mathcal L}_\mu$ and ${\mathcal R}_\mu$,
we define the external axial-vector and vector fields as
${\mathcal A}_\mu = ( {\mathcal R}_\mu - {\mathcal L}_\mu )/2$ and
${\mathcal V}_\mu = ( {\mathcal R}_\mu + {\mathcal L}_\mu )/2$.

In the HLS theory it is possible to perform the derivative
expansion systematically~\cite{Georgi,Tanabashi:1993sr,HY:WM,HY:PR}.
In the chiral perturbation theory (ChPT) with  HLS,
the vector meson mass is to be considered as small compared with
the chiral symmetry breaking scale $\Lambda_\chi$,
by assigning ${\mathcal O}(p)$ to the HLS gauge coupling,
$g \sim {\mathcal O}(p)$~\cite{Georgi,Tanabashi:1993sr}.
(For details of the ChPT with HLS, see Ref.~\cite{HY:PR}.)
The leading order Lagrangian with
Lorentz non-invariance can be written as~\cite{VM:dense}
\begin{eqnarray}
&&
{\mathcal L}
=
\biggl[
  (F_{\pi}^t)^2 u_\mu u_\nu
  {}+ F_{\pi}^t F_{\pi}^s
    \left( g_{\mu\nu} - u_\mu u_\nu \right)
\biggr]
\nonumber\\
&&\hspace*{1cm}\times
\mbox{tr}
\left[
  \hat{\alpha}_\perp^\mu \hat{\alpha}_\perp^\nu
\right]
\nonumber\\
&&{}+
\biggl[
  (F_{\sigma}^t)^2 u_\mu u_\nu
  {}+  F_{\sigma}^t F_{\sigma}^s
    \left( g_{\mu\nu} - u_\mu u_\nu \right)
\biggr]
\nonumber\\
&&\hspace*{1cm}\times
\mbox{tr}
\left[
  \hat{\alpha}_\parallel^\mu \hat{\alpha}_\parallel^\nu
\right]
\nonumber\\
&&
{} +
\Biggl[
  - \frac{1}{ g_{L}^2 } \, u_\mu u_\alpha g_{\nu\beta}
\nonumber\\
&&\hspace*{1cm}
  {}- \frac{1}{ 2 g_{T}^2 }
  \left(
    g_{\mu\alpha} g_{\nu\beta}
   - 2 u_\mu u_\alpha g_{\nu\beta}
  \right)
\Biggr]
\nonumber\\
&&\hspace*{1cm}\times
\mbox{tr}
\left[ V^{\mu\nu} V^{\alpha\beta} \right]
\ ,
\label{Lag}
\end{eqnarray}
where
\begin{equation}
 \hat{\alpha}_{\perp,\parallel }^{\mu}
 = \frac{1}{2i}\bigl[ D^\mu\xi_R \cdot \xi_R^{\dagger}
                 {}\mp  D^\mu\xi_L \cdot \xi_L^{\dagger}
                   \bigr].
\end{equation}
Here $F_{\pi}^s$ denote the spatial pion decay constant and similarly
$F_{\sigma}^s$ for the $\sigma$. The rest frame
of the medium is specified by $u^\mu = (1,\vec{0})$ and
$V_{\mu\nu}$ is the field strength of $V_\mu$. $g_{L}$ and $g_{T}$
correspond in medium to the HLS gauge coupling $g$. The parametric
$\pi$ and $\sigma$ velocities are defined by~\cite{Pisarski:1996mt}
\begin{equation}
 V_\pi^2 = {F_\pi^s}/{F_\pi^t}, \qquad
 V_\sigma^2 = {F_\sigma^s}/{F_\sigma^t}.
\end{equation}

Now we approach the critical point of chiral symmetry restoration,
which is characterized by the equality between the axial-vector
and vector current correlators in QCD, $G_A - G_V \to 0$ for $T
\to T_c$. This should hold even in the EFT side. In
Ref.~\cite{VM:dense}, it was shown that they are satisfied for any
values of $p_0$ and $\bar{p}$ around the matching scale only if
the following conditions are met: $(g_{L,{\rm bare}}, g_{T,{\rm
bare}}, a_{\rm bare}^t, a_{\rm bare}^s) \to (0,0,1,1)$ for $T \to
T_c$. This implies that at the bare level the longitudinal mode of
the vector meson becomes the real NG boson and couples to the
vector current correlator, while the transverse mode decouples. As
shown in Ref.~\cite{sasaki}, $(g_L, a^t, a^s) = (0,1,1)$ is a
fixed point of the RGEs and satisfied at any energy scale. Thus
the VM condition is given by
\begin{eqnarray}
 (g_L, a^t, a^s) \to (0,1,1) \quad \mbox{for}\quad T \to T_c.
\label{evm}
\end{eqnarray}
The vector meson mass is never generated at the critical
temperature since the quantum correction to $M_\rho^2$ is
proportional to $g_L^2$. Because of $g_L \to 0$, the transverse
vector meson at the critical point, at any energy scale, decouples
from the vector current correlator. The VM condition for $a^t$ and
$a^s$ leads to the equality between the $\pi$ and $\sigma$ (i.e.,
longitudinal vector meson) velocities:
\begin{eqnarray}
 \bigl( V_\pi / V_\sigma \bigr)^4
 = \bigl( F_\pi^s F_\sigma^t / F_\sigma^s F_\pi^t \bigr)^2
 = a^t / a^s
 \stackrel{T \to T_c}{\to} 1.
\label{vp=vs}
\end{eqnarray}
This can be easily understood from the point of view of the VM
since the longitudinal vector meson becomes the chiral partner of
pion. We note that this condition $V_\sigma = V_\pi$ holds
independently of the value of the bare pion velocity which is to
be determined through the Wilsonian matching.


\section{Pion velocity near the critical temperature}
One possible way to determine the bare parameters is the Wilsonian
matching which is done by matching the axial-vector and vector
current correlators derived from the HLS with those by the
operator product expansion (OPE) in QCD at the matching scale
$\Lambda$~\cite{HY:WM}. The Wilsonian matching leads to the
following conditions on the bare pion decay
constants~\cite{HKRS:pv}:
 \begin{eqnarray}
 &&
  \frac{F_{\pi,{\rm bare}}^t F_{\pi,{\rm bare}}^s}{\Lambda^2}
 = \frac{1}{8\pi^2}\Biggl[ \Bigl( 1 + \frac{\alpha_s}{\pi} \Bigr)
\nonumber\\
&&\quad{}+
   \frac{2\pi^2}{3}\frac{\big\langle \frac{\alpha_s}{\pi}G^2
   \big\rangle_T }{\Lambda^4}
   {}+ \pi^3 \frac{1408}{27}\frac{\alpha_s
    \langle \bar{q}q \rangle_T^2}{\Lambda^6}  \Biggr]
\nonumber\\
 &&\quad
  {}+ \frac{\pi^2}{15}\frac{T^4}{\Lambda^4}A_{4,2}^\pi
  {}- \frac{16\pi^4}{21}\frac{T^6}{\Lambda^6}A_{6,4}^\pi
 \,\,\equiv G_0,
\nonumber\\
 &&
  \frac{F_{\pi,{\rm bare}}^t F_{\pi,{\rm bare}}^s
 (1 - V_{\rm bare}^2)}{\Lambda^2}
  = \frac{32}{105}\pi^4\frac{T^6}{\Lambda^6} A_{6,4}^\pi,
\nonumber\\
\end{eqnarray}
where we use the dilute pion-gas approximation in order to
evaluate the matrix element $\langle {\mathcal O} \rangle_T$~\cite{hkl}
in the low temperature region.
{}From these conditions, we obtain the following matching condition to
determine the deviation of the bare pion velocity from the speed of
light in the low temperature region:
\begin{equation}
 \delta_{\rm bare}\equiv 1 - V_{\pi,{\rm bare}}^2
 = \frac{1}{G_0}
   \frac{32}{105}\pi^4\frac{T^6}{\Lambda^6} A_4^{\pi} .
\label{deviation-rho}
\end{equation}
This implies that the intrinsic
temperature dependence starts from the ${\mathcal O}(T^6)$
contribution.

As is
discussed in Ref.~\cite{HKRS:pv}, we should in principle evaluate
the matrix elements in terms of QCD variables only in order for
performing the Wilsonian matching, which is as yet unavailable from
model-independent QCD calculations.  Therefore, we make an
estimation by extending the dilute gas approximation adopted in
the QCD sum rule analysis in low temperature region to the
critical temperature including all the light degrees of
freedom expected in the VM. In the HLS/VM theory, both the
longitudinal and transverse vector mesons become massless at the
critical temperature.
At the critical point, the longitudinal vector meson
couples to the vector current
whereas the transverse vector mesons decouple from the theory.
Thus we assume that thermal
fluctuations of the system are dominated near $T_c$ not only by
the pions but also by the longitudinal vector mesons.
We evaluate the thermal matrix elements of the non-scalar operators
in the OPE, by extending the thermal pion gas approximation employed
in Ref.~\cite{hkl} to the longitudinal vector mesons that figure
in our approach.

This is feasible since at the critical
temperature, we expect the equality $A_4^\rho(T_c) = A_4^\pi(T_c)$
to hold as the massless longitudinal vector meson is the chiral
partner of the pion in the VM.
It should be noted that, although we use the dilute gas
approximation, the treatment here is already beyond the
low-temperature approximation because
the contribution from vector meson is negligible in the
low-temperature region. Since we treat the pion as a massless
particle in the present analysis, it is reasonable to take
$A_4^\pi(T) \simeq A_4^\pi(T=0)$. We therefore use
\begin{equation}
 A_4^\rho(T) \simeq A_4^{\pi}(T) \simeq A_4^\pi(T=0)
 \quad \mbox{for}\quad T \simeq T_c.
\label{matrix Tc}
\end{equation}
Therefore from Eq.~(\ref{deviation-rho}), we obtain the deviation
$\delta_{\rm bare}$ as
\begin{equation}
 \delta_{\rm bare} = 1 - V_{\pi,{\rm bare}}^2
 = \frac{1}{G_0}
   \frac{32}{105}\pi^4\frac{T^6}{\Lambda^6}
  \Bigl[ A_4^{\pi}+  A_4^{\rho} \Bigr].
\label{deviation-pi-rho}
\end{equation}
This is the matching condition to be used for determining the
value of the bare pion velocity near the critical temperature.
Let us make a rough estimate of $\delta_{\rm bare}$.
For the range of matching scale
$(\Lambda = 0.8 - 1.1\, \mbox{GeV})$, that of QCD scale
$(\Lambda_{QCD} = 0.30 - 0.45\, \mbox{GeV})$ and critical
temperature $(T_c = 0.15 - 0.20\, \mbox{GeV})$, we get
\begin{equation}
 \delta_{\rm bare}(T_c) = 0.0061 - 0.29\,.
\end{equation}
Thus we obtain the $bare$ pion velocity as
$V_{\pi,{\rm bare}}(T_c) = 0.83 - 0.99\,$.

We next consider the quantum and hadronic thermal corrections to
the parametric pion velocity. It was proven in Ref.~\cite{sasaki}
that the pion velocity is protected from renormalization by the
VM. In the following, we show that this can be understood in terms
of chiral partners: Away from $T_c$, the pion velocity receives
hadronic thermal correction of the form~\cite{HS:VVD}:
\begin{eqnarray}
 v_\pi^2 (T) &\simeq&
 V_\pi^2 - N_f \frac{2\pi^2}{15}\frac{T^4}{(F_\pi^t)^2 M_\rho^2}
\nonumber\\
&& \mbox{for} \quad T < T_c, \label{low T}
\end{eqnarray}
where the contribution of the massive $\sigma$ (i.e., the
longitudinal mode of massive vector meson) is suppressed by the
Boltzmann factor $\exp [-M_\rho / T]$, and then only the pion loop
contributes to the pion velocity. On the other hand, when we
approach the critical temperature, the vector meson mass goes to
zero due to the VM. Thus $\exp [-M_\rho / T]$ is no longer the
suppression factor. As a result, the hadronic correction in the
pion velocity is absent due to the exact cancelation between the
contribution of pion and that of its chiral partner $\sigma$.
Similarly the quantum correction generated from the pion loop is
exactly canceled by that from the $\sigma$ loop. Accordingly we
conclude
\begin{equation}
 v_\pi(T) = V_{\pi,{\rm bare}}(T)
 \qquad \mbox{for}\quad T \to T_c,
\label{phys=bare}
\end{equation}
i.e., {\it the pion velocity in the limit $T \to T_c$ receives
neither hadronic nor quantum corrections due to the protection by
the VM.} This implies that $(g_L,a^t,a^s,V_\pi) =
(0,1,1,\mbox{any})$ forms a fixed line for four RGEs of $g_L, a^t,
a^s$ and $V_\pi$. When a point on this fixed line is selected
through the matching procedure as explained in
Ref.~\cite{HKRS:pv}, that is to say when the value of $V_{\pi,{\rm
bare}}$ is fixed, the present result implies that the point does
not move in a subspace of the parameters. Approaching the chiral
symmetry restoration point, the physical pion velocity itself will
flow into the fixed point. Finally thanks to the
non-renormalization property, i.e., $v_\pi (T_c)=V_{\pi, {\rm
bare}}(T_c)$ given in Eq.~(\ref{phys=bare}), we arrive at the
physical pion velocity at the chiral restoration:
\begin{equation}
 v_{\pi}(T_c) = 0.83 - 0.99\,,\label{ours}
\end{equation}
close to the speed of light.


\section{Conclusion}
In this note, we studied, using the ChPT with HLS/VM, the pion
velocity near the critical temperature. We exploited the
non-renormalization property of the pion velocity to assure that
it suffices to compute the $bare$ pion velocity at the matching
scale to arrive at the $physical$ pion velocity at the chiral
restoration temperature. We derived the matching condition on the
$bare$ pion velocity and found that the pion velocity near $T_c$
is close to the speed of light, $v_\pi^{\rm (VM)} (T) = 0.83 -
0.99\,$ and definitely far from the zero velocity. This is in a
stark contrast to the result obtained from the chiral theory
~\cite{SS}, wherein only the pion figures as the relevant degree
of freedom near $T_c$, namely, $v_\pi (T_c)=0$. The drastic
difference between the two approaches is not difficult to
understand. In the HLS/VM approach, the $\rho$ meson becomes light
as $T_c$ is approached from below and plays as important a role as
the pion does. The effect of the massless vector meson cannot be
approximated in chiral models by local operators in pseudoscalar
fields.

By fitting the pion spectra observed by STAR at RHIC in terms of
an optical potential that incorporates the dispersion relation of
low-energy pions in nuclear matter, Cramer et al. deduced the
in-medium pion velocity $v_\pi (T) = 0.65$. The authors
interpreted this result as an evidence for the pions being emitted
from the chiral-symmetry restored phase. No error bars have been
assigned to this value, so it is difficult to make a clear-cut
assessment of what that value implies. It seems however difficult
to identify it with what has been predicted by sigma models,
namely, $v_\pi=0$. On the other hand, given that our prediction
(\ref{ours}) based on HLS/VM is for the chiral limit, it seems
reasonable to expect that the account of the explicit chiral
symmetry breaking by quark masses would lower the velocity from
(\ref{ours}), making it closer to the observed value,
$v_\pi=0.65$. Whether or not it signals the chirally restored
phase as interpreted in \cite{Cramer, Wilczek} is not clear.
However as argued in ~\cite{bgr}, there is an indication for a
massless pion (in the same multplet with a scalar $\sigma$) {\it
just above} $T_c$ with its velocity close to 1, it seems logical
that $v_\pi$ stays near 1 -- rather than near zero -- as $T_c$ is
approached from below as well as from above.


\section*{Acknowledgments}
We are grateful for discussions with Gerry Brown, Youngman Kim,
Koichi Yamawaki.
This work is supported in part by
the JSPS Grant-in-Aid for Scientific Research (c) (2) 16540241,
and by
the 21st Century COE
Program of Nagoya University provided by Japan Society for the
Promotion of Science (15COEG01).


\end{document}